\documentclass[12pt]{iopart}
\usepackage{graphicx}
\graphicspath{{./images/}}
\usepackage{amsfonts}
\usepackage{amssymb}
\usepackage{amsthm}
\usepackage{color}
\usepackage{soul}
\theoremstyle{definition} 
\theoremstyle{plain} 
\theoremstyle{remark} 
\usepackage{graphicx}

\begin{document}
\title[]{Phase transitions in integer linear problems.}
\author{S. Colabrese$^1$, D. De Martino$^{2}$, L. Leuzzi$^{3,4}$ and E. Marinari$^{4,3,5}$}
\address{
  $^1$ Dip. di Fisica, Univ. Tor Vergata, I-00133 Rome, Italy \\
  $^2$ Institute of Science and Technology Austria (IST Austria),
  Am Campus 1, Klosterneuburg A-3400, Austria \\
  $^3$ Soft and Living Matter Lab., Rome Unit of CNR-NANOTEC,
  Institute of Nanotechnology, National Research Council of Italy,
  P.le A. Moro 5, I-00185, Rome, Italy \\
  $^4$ Dipartimento di Fisica, Sapienza Universit\'{a} di Roma,
  P.le A.Moro 2, 00815, Rome, Italy \\
  $^5$ INFN, Sezione di Roma 1.}
\begin{abstract}
The resolution of linear system with positive integer variables is a
basic yet difficult computational problem with many applications.
We consider sparse uncorrelated random systems parametrised
by the density $c$ and the ratio $\alpha=N/M$ between number of
variables $N$ and number of constraints $M$. By means of ensemble
calculations we show that the space of feasible solutions endows a
Van-Der-Waals phase diagram in the plane ($c$, $\alpha$).  We give
numerical evidence that the associated computational problems become
more difficult across the critical point and in particular in the
coexistence region.
\end{abstract}

\maketitle

\section{Introduction}

The use of concepts and techniques from statistical mechanics to
analyze hard combinatorial problems has been very
fruitful\cite{martin2001statistical, mezard2009information}, with
examples that range from the paradigmatic
K-sat\cite{monasson1997statistical,leuzzi2001the,crisanti2002the,mezard2002analytic}
to the traveling salesman problem\cite{kirkpatrick1985configuration}
and number partitioning\cite{mertens1998phase} to cite few.

Generically, upon defining an ensemble of random instances of the
problem described by some parameters we can analyze how the
solutions, and their structure, depend on these parameters.  A typical
example is a system with constraints connecting different
variables. One significant parameter will, then, be the average
connectivity of a single variable and another one the ratio between
the number of variables and the number of constraints.  If we are able
to assess the statistics of the number of solutions $N_{sol}$ and
compute the average $\langle \log N_{sol} \rangle$ over different
instances of the problem, we can adopt this {\em solution entropy},
possibly extensive, as a thermodynamic potential and look at
singularities in the derivatives with respect to the parameters that
can be interpreted as phase transitions among different
phases \cite{hartmann2006phase, mertens2002computational}.

It turns out that, typically, we can distinguish a region in the
parameter space in which solutions to the problem can always be found
- the SAT phase - from a region in which no solution can be found -
the UNSAT phase. The associated decision problem,
i.e. determining whether a solution exists or not, can be
efficiently solved in regions deeply within each phase whereas it
becomes effectively hard on the boundary between them
\cite{monasson1999determining}.
 
A fundamental problem in combinatorics deals with the resolution of
linear systems with non negative integer variables.  The latter
integrality constraint makes this issue computationally hard but still
it inherits some useful results from linear and convex
algebra\cite{schrijver1998theory}. Several kinds of problems can be
assessed: decide if the system has non-trivial solutions (decision
problem), describe the space by exhibiting a so-called Hilbert
basis\cite{henk1996hilbert}, count the lattice points inside a convex
region\cite{barvinok2008integer} or maximize a linear function over
integer vectors in a polyhedron (integer linear
programming\cite{lenstra1983integer}).  The latter problem is
particularly interesting since it admits a straightforward
approximation upon relaxing the integrality constraint and thus
defining a linear programming problem, that is efficiently solvable in
polynomial time and is connected to an important dual 
problem\cite{schrijver1998theory}. A fundamental strategy in
combinatorial optimization thus consists in formulating problems as
integer programs.  Apart from classical applications in operations
research and planning, integer-linear modeling is massively used in the
field of constraint-based models of metabolic
networks, in which the resolution of such systems is required for the
computation of extreme pathways\cite{schilling2000theory}, elementary
flux modules\cite{schuster1994elementary}, conserved
moieties\cite{schuster1991determining, de2014identifying} and
thermodynamically unfeasible cycles\cite{muller2012thermodynamic,
  de2013counting}.  We do point out, further, the recent surge of
interest in statistical physics of the Gardner
problem\cite{gardner1988space} of calculating the volume of the space
of interactions of neural network models for given patterns, as it has
been shown that this can be considered as a mean field model of
jamming\cite{franz2016simplest} of hard objects, that - in turn - can
be connected to the resolution of integer linear systems.  Indeed, it
has been recently shown that the feasibility of the constraints
defining such spaces is ruled by duality theorems defining dual spaces
representing unfeasible patterns, that are represented by integer
linear systems of equations\cite{de2016dual}.

Inspired by the aforementioned statistical mechanical approaches,
we analyze here the solution spaces of large random sparse
integer linear systems. In Sec. \ref{model} we define the model and
the ensemble, in Sec. \ref{results} we present the thermodynamical
behavior of these problems and their phase diagram obtained by means
of annealed and of quenched replica calculations.  In
Sec. \ref{single}, we show how this framework can be used to analyze
algorithms apt to solve integer programs and, finally, draw our
conclusions.

\section{The model}
\label{model}
We consider a linear system of $M$ equations in $N$ unknowns $n_j$:
\begin{equation}
  \label{eq:system}
  \sum_{j = 1}^{N} \xi_{j,\mu} n_j = 0  \qquad  \mu = 1, \cdots, M\; .
\end{equation}
We will consider both the case where the variables $n_i$ are
non-negative integers and the case where they are Boolean variables,
i.e. they can only take the values $0$ and $1$.  We will consider an
ensemble of random systems where the coefficients $\xi_{j,\mu}$ are
independent, identically distributed random variables:
\begin{equation}
  \label{eq:ensemble}
  \xi_{j,\mu} = \left\{ \begin{array}{rlc}
  0 & \mbox{with probability } & 1-p\;,\\
  1 & \mbox{with probability } & p/2\;,\\
  -1 & \mbox{with probability } & p/2\;.
\end{array} \right.
\end{equation}
We study systems where $p = \frac{c}{N}$ (with $c$ a intensive
constant). These systems are sparse, 
with a Poissonian structure, where $c$ is the average
number of variables per equation.  We will analyze large systems, where
$N\to \infty$, $M \to \infty$, keeping $\alpha \equiv\frac{N}{M}$
fixed.

We can map the solutions of system (\ref{eq:system}) onto
ground states of a model whose Hamiltonian has the form
\begin{equation}
  \label{eq:statmech}
  H = \sum_\mu \left(\sum_j \xi_{j,\mu} n_i\right)^2
    = \sum_{j,k} J_{j,k} n_j n_k \;,
\end{equation}
where
\begin{equation}
  \label{eq:couplings}
  J_{j,k} = \sum_\mu \xi_{j\mu} \xi_{k \mu}\;.
\end{equation}
It is interesting to
consider the more general problem of calculating the partition function
\begin{equation}
  \label{eq:zeta}
  Z = \sum_{\{n_j\}} e^{-\beta H}\;.
\end{equation}
For example computing $Z$ in the limit $\beta \to \infty$ gives the
number of solutions of (\ref{eq:system}).

The case where $n_j=0,1$ will turn out to be different from the
general case where the $n_j$ are semi-positive definite integer
variables (always with $\xi$ coefficients that can take the three
values $0$, $+1$ and $-1$).

\section{The disorder average}
\label{results}

\subsection{The annealed approximation for $n_i=0,1$, finite T.}
We will start by studying the $n_i=0,1$ model in the annealed approximation,
i.e. by averaging the partition function (and not its logarithm, as we will
do in the quenched approach) over the disorder
(\ref{eq:ensemble}). We denote by an over-line the average over the
disorder. We have that
\begin{eqnarray}
\label{eq:anneal}
\nonumber
\overline{Z}
= \sum_{\{n_j\}} \overline{e^{-\beta H}} \\
= \sum_{\{n_j\}} \prod_\mu \int dm_\mu\; e^{-\beta m_\mu^2}\;
\overline{\delta(m_\mu - \sum_j \xi_{j,\mu} n_j)} \nonumber\\
=\left(\frac{1}{2\pi}\right)^{M}\sum_{\{n_j\}}
\int \int
\left(\prod_{\mu=1,M}d\lambda_\mu\right)
\left(\prod_{\mu=1,M}dm_\mu\right)
e^{-\beta |\vec{m}|^2 + i \vec{\lambda} \cdot \vec{m}} \prod_{j,\mu}
\overline{e^{-i n_j \lambda_\mu \xi_{j,\mu} }} \nonumber\\
=\left(\frac{1}{4\pi\beta}\right)^{\frac{M}{2}}
\int \left(\prod_{\mu=1,M}d\lambda_\mu\right)
e^{-\frac{1}{4\beta} |\vec{\lambda}|^2} \sum_{\{n_j\}}
\prod_{j,\mu}(1-p + p \cos(\lambda_\mu n_j)) \nonumber \\
=\bigg(\frac{1}{4\pi\beta}\bigg)^{\frac{M}{2}}
\int \left(\prod_{\mu=1,M}d\lambda_\mu\right)
e^{-\frac{1}{4\beta} |\vec{\lambda}|^2}
\left(1+ \prod_\mu \left(1-p + p \cos\left(\lambda_\mu\right)\right)\right)^N\;,
\nonumber 
\end{eqnarray}
and, by developing the binomial,
\begin{equation}
\overline{Z}
= \sum_{k=0}^{N} {N \choose k} (1-p)^{kM} \left(
\frac{1}{\sqrt{4\pi\beta}}\int
d\lambda\;   e^{-\frac{1}{4\beta} \lambda^2} \left(1 + \frac{p}{1-p}
\cos\lambda\right)^k\right)^M\;.
\end{equation}
\begin{figure}[ht]
\centering
\includegraphics*[width=.45\textwidth,angle=0]{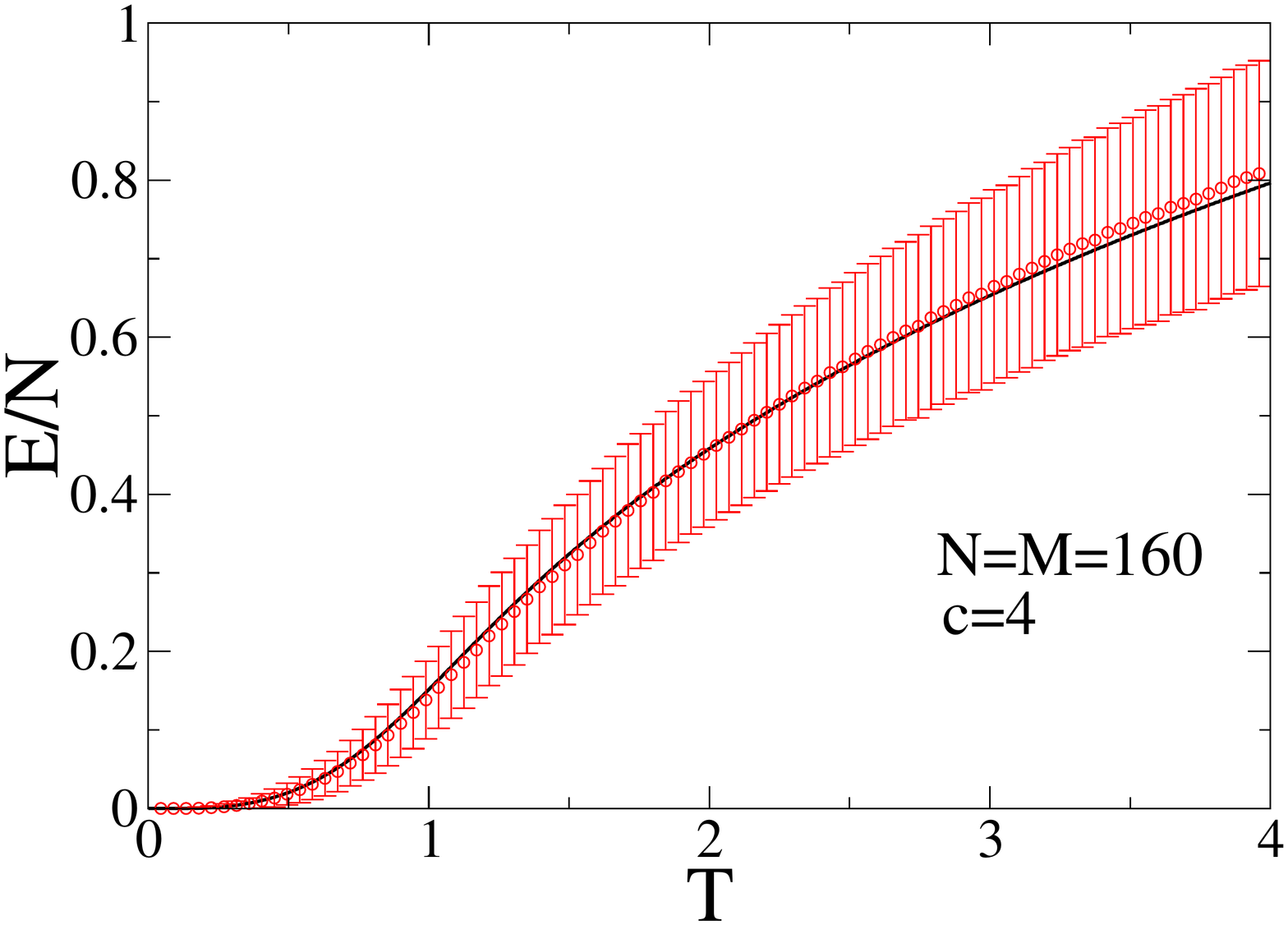}
\includegraphics*[width=.45\textwidth,angle=0]{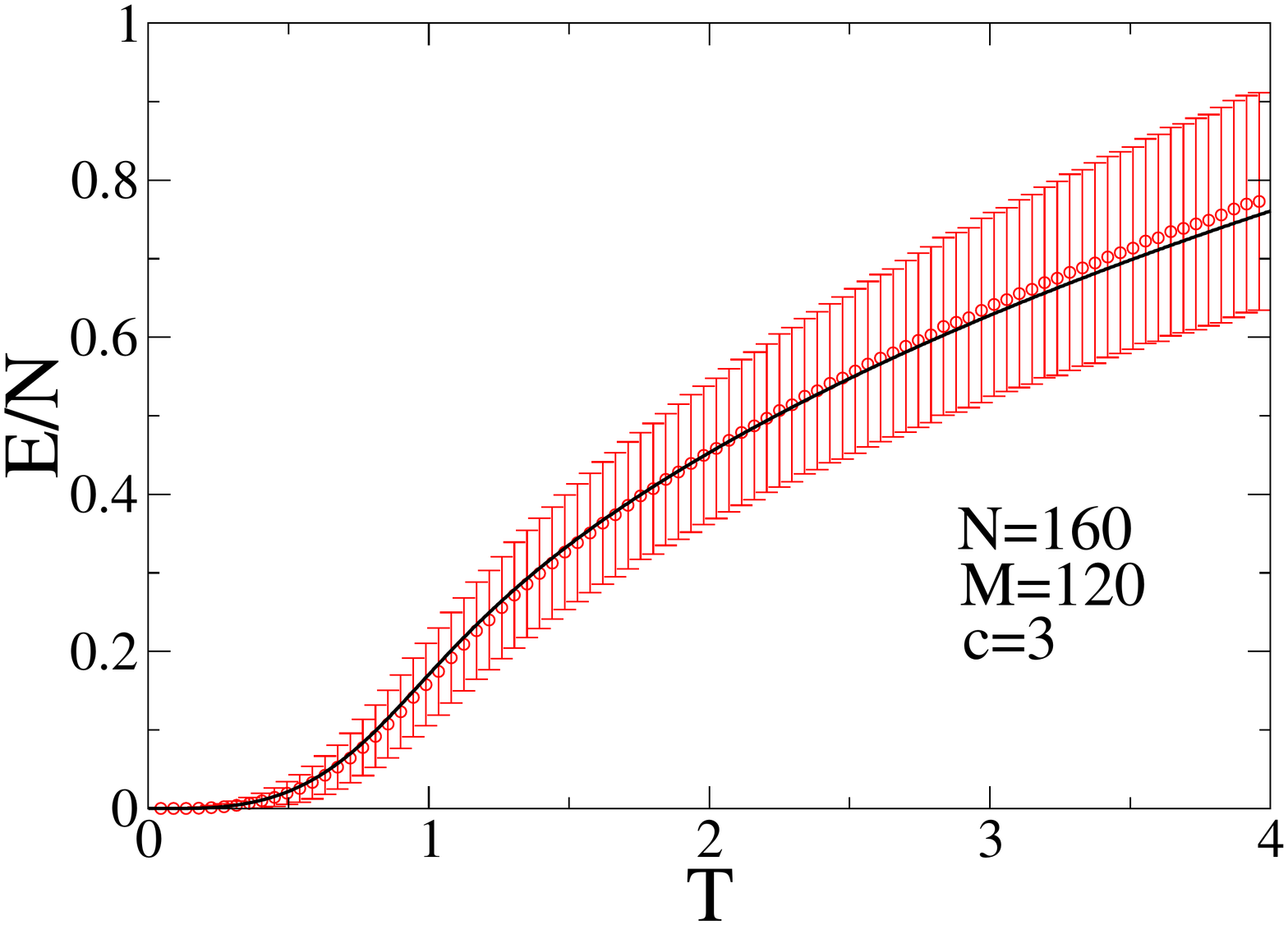}
\includegraphics*[width=.45\textwidth,angle=0]{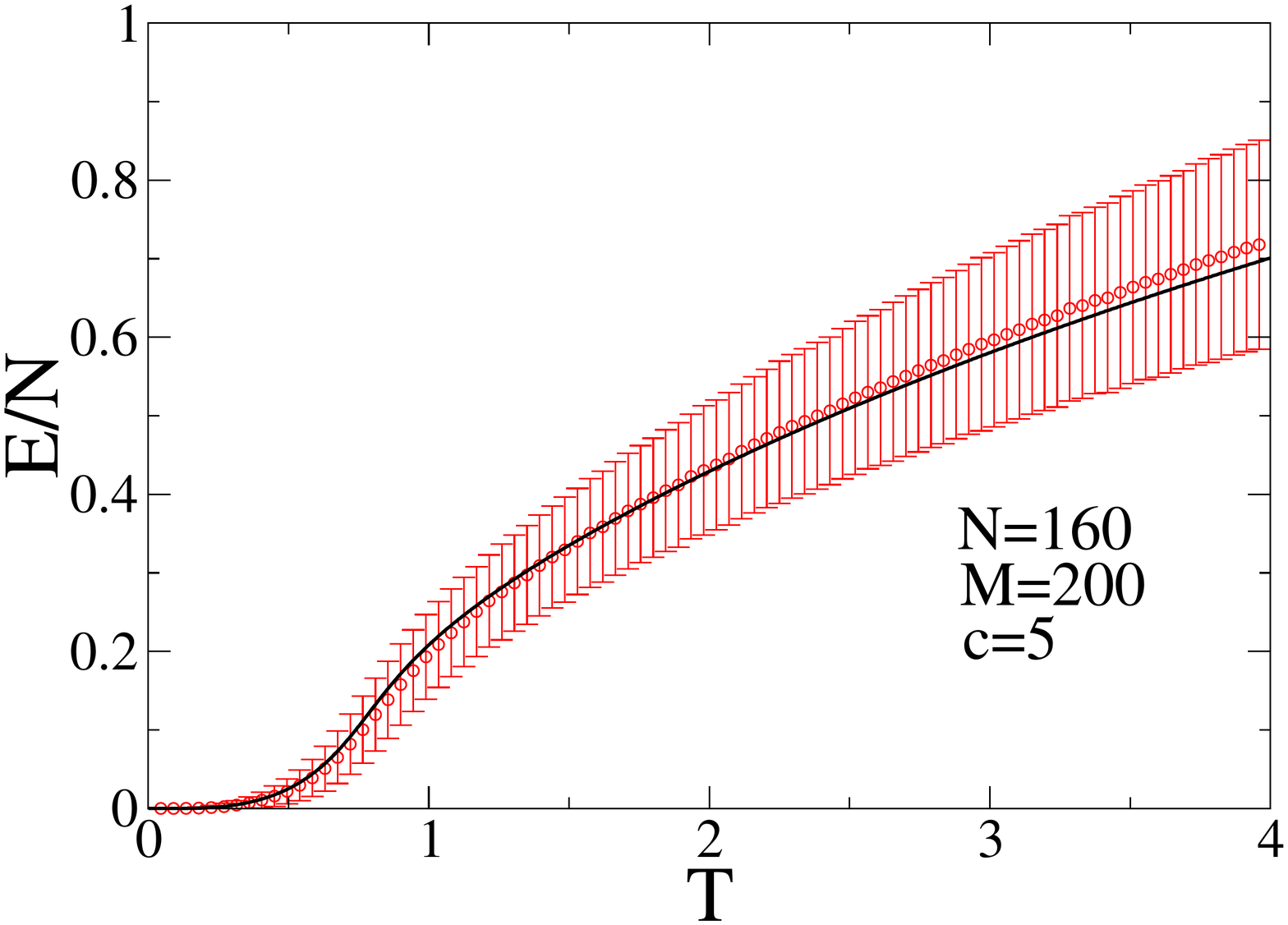}
\caption{Energy curves $E(T)$ from annealed calculations (lines) and measured in Monte Carlo simulations over $100$ instances (circles). $N=160$ Top left: $M=160$, Top right $M=120$, Bottom $M=200$. $c\cdot \alpha =4$.}
\label{fig:fig1}
\end{figure}
Since $x\equiv \frac{p}{1-p}$ is small, we can set
$(1+x\cos\lambda)^k \simeq e^{k x \cos \lambda}$ and use the relation
\begin{equation}
  e^{kx \cos \lambda} = I_0(kx) + 2 \sum_{n=1}^\infty I_n(kx) \cos (n\lambda)\;,
\end{equation}
where the $I_n$ are the modified Bessel functions of the first kind
of order $n$ (\cite{abramowitz1964handbook}).

Now the Gaussian integral can be evaluated by noticing that 
\begin{equation}
  \frac{1}{\sqrt{4\pi\beta}} \int d\lambda\;
  e^{-\frac{\lambda^2}{4\beta} } \cos (n\lambda) = e^{-\beta n^2}\;,
\end{equation}
and it gives
\begin{equation}
  \overline{Z}
  \simeq \sum_{k = 0}^{N} {N \choose k} (1-p)^{kM}
  (I_0(kx) + 2 \sum_{n=1}^\infty I_n(kx)
  e^{-\beta n^2})^M\;.
\end{equation}
By using the Stirling approximation for the binomial coefficients and
the fact that $p$ is small (that allows to write $-p$ for $\log(1-p)$)
we write
\begin{equation}
  \overline{Z} \equiv \sum_{k = 0}^{N} e^{N f_{\alpha,\beta,c}(k/N)}\;,
\end{equation}
that defines
\begin{equation}
  f_{\alpha,\beta,c}(r) = -r \log r -(1-r) \log (1-r) -\frac{c}{\alpha} r
  + \frac{1}{\alpha} \log(G_c(r))\;,
\end{equation}
where $r \equiv \frac{k}{N}$ and
\begin{equation}
  G_c(r) \equiv I_0(cr) + 2\sum_n I_n(cr) e^{-\beta n^2}\:.
\end{equation}
The use of the Stirling approximation is justified when $M$ and $N$
are large with $c$ finite (i.e. our approach is valid for diluted
models). We have also verified this fact numerically, by exact
enumeration.  We can now evaluate $\overline{Z}$ in our approximation
by a saddle point approximation. We have to find the maxima of $f$ as
a function of $r$ for $r \in [0,1]$ ($r$ emerges here as a natural
order parameter). 

The order parameter $r$ that emerges from the saddle point calculation
is the average number of active variables per solution of system
(\ref{eq:system}). Its asymptotic value for $N\gg M$, where every vector
variable is a solution, is $\frac12$.
 
 By deriving $f_{\alpha,\beta,c}(r)$ with respect to
$r$ we obtain a self consistent equation for its stationary points
\begin{equation}
r  = (1-r) e^{\frac{1}{\alpha}(G'/G -c)}\;,
\end{equation}
that can be solved numerically for each value of $c,\alpha,\beta$ in
order to obtain thermodynamic quantities like the free energy ($F =
-\frac{N}{\beta} f$), the energy ($ E = N \frac{d f}{d\beta}$) and the
entropy ($S = \beta (E-F)$). In Fig. \ref{fig:fig1} we show
the energy density obtained from these annealed analytic
computation for different choices of $\alpha$, $\beta$ and $c$ and from a
Monte Carlo simulations (of the original, quenched theory).  No
unexpected effects are seen, and the agreement is very reasonable.

\begin{figure}[h]
\centering
\includegraphics*[width=.45\textwidth,angle=0]{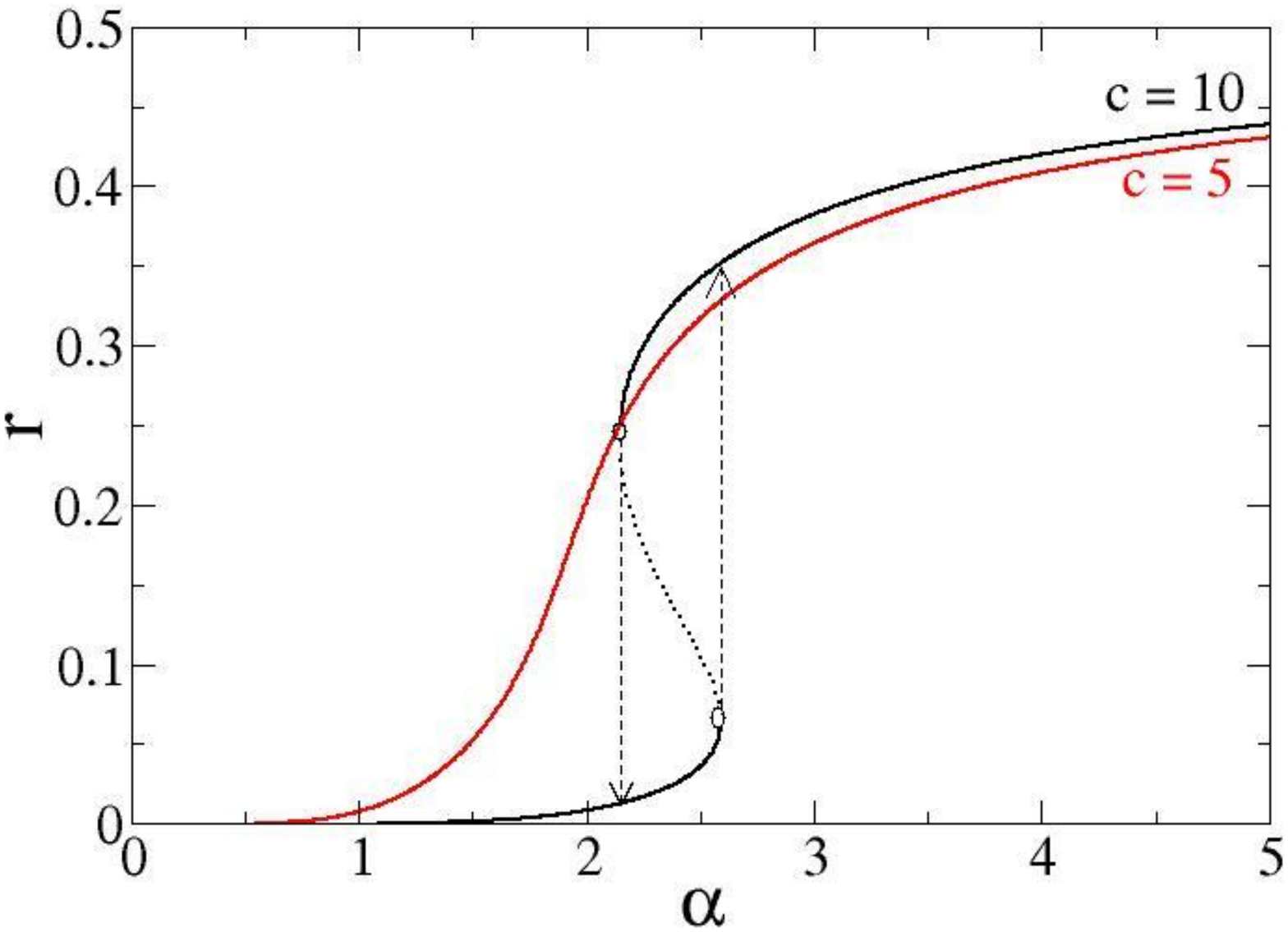}
\includegraphics*[width=.45\textwidth,angle=0]{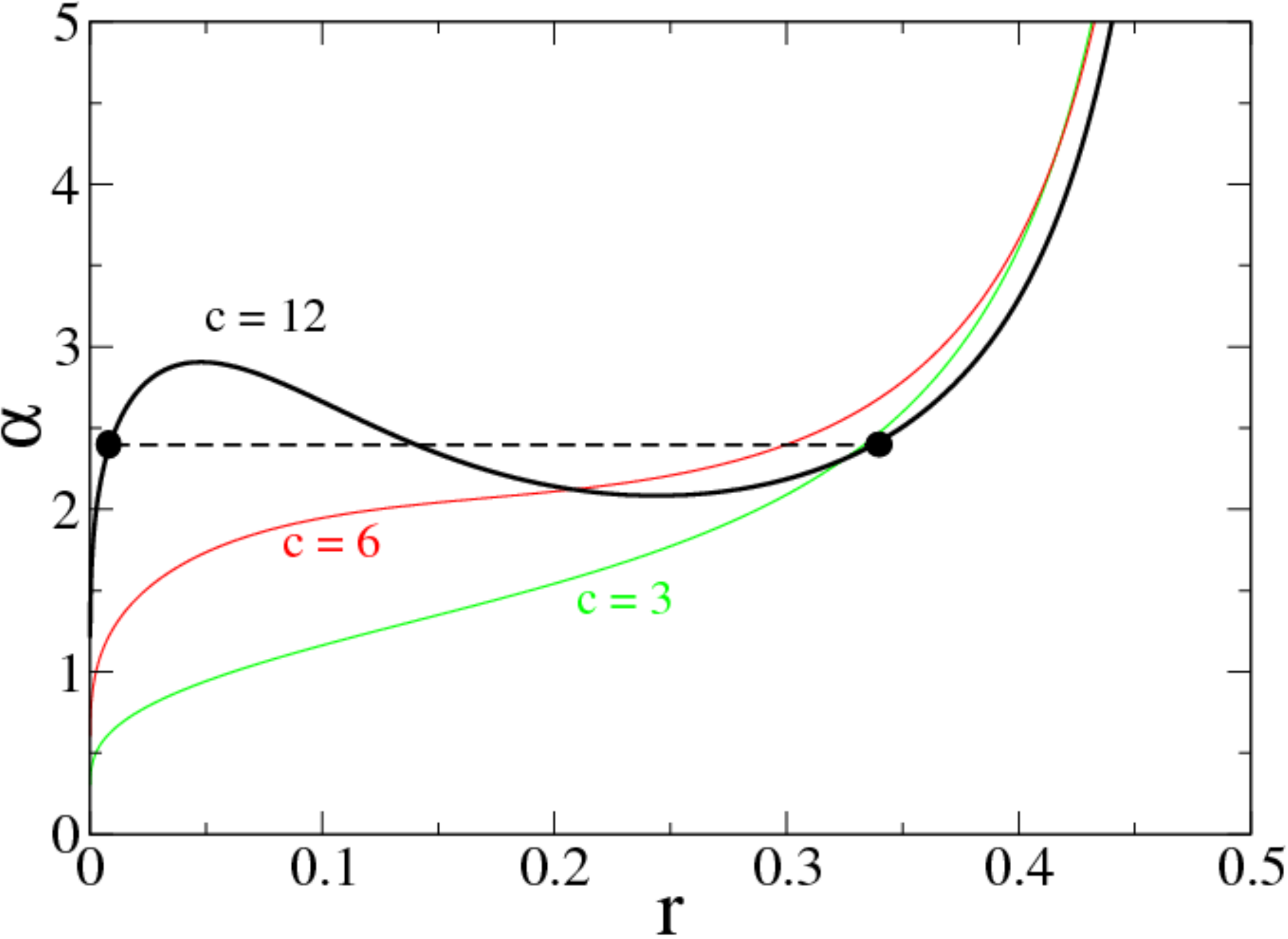}
\includegraphics*[width=.7\textwidth,angle=0]{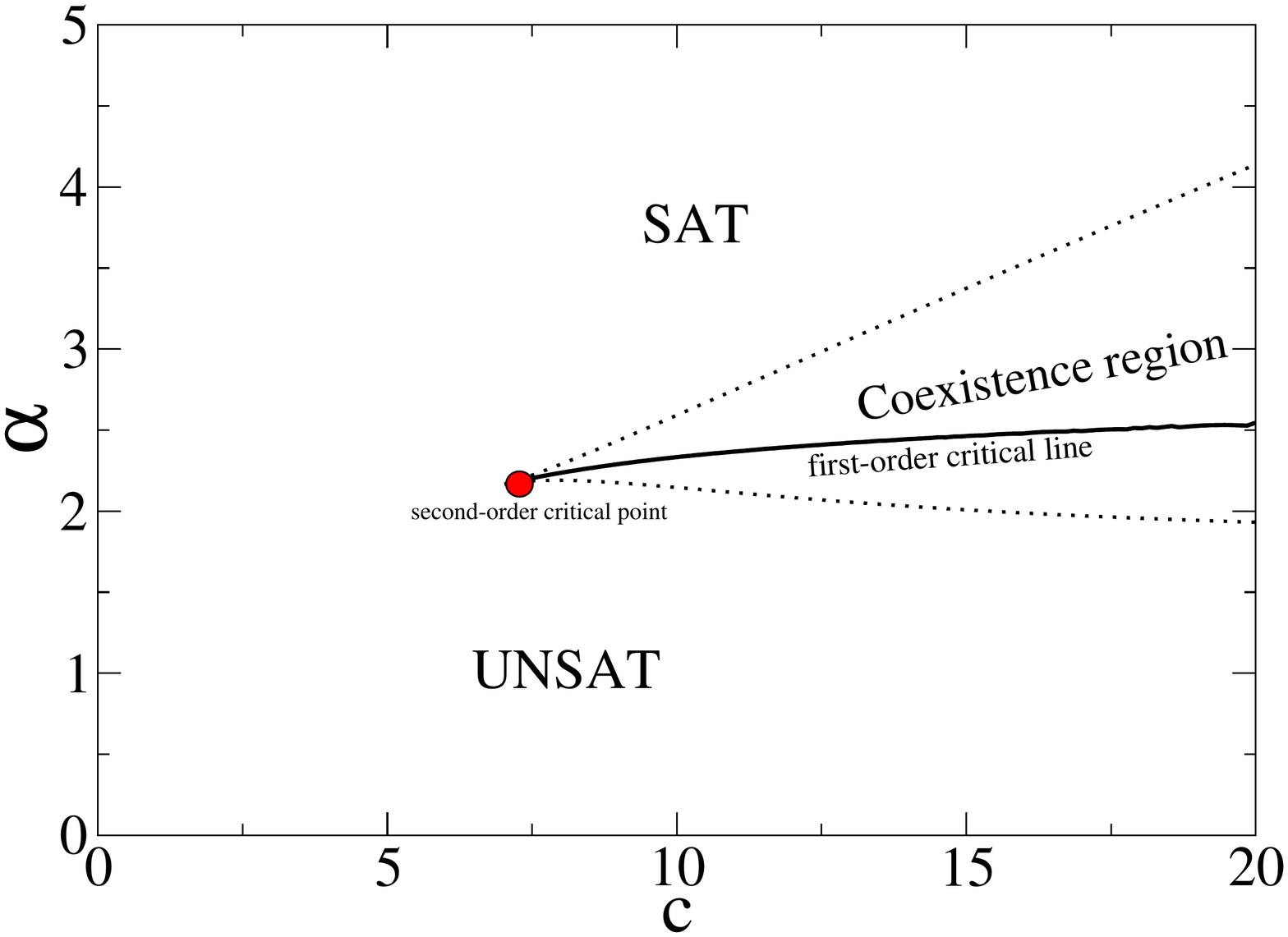}
\caption{Top Left: the curves $r(\alpha)$ for $c=5$ and $c=10$
  respectively. For $c=10$ there are two stable solutions in a range
  of $\alpha$ (first order transition). Top Right: The curves $\alpha(r)$
  for $c=3,6,12$. For $c=12$ we show the Maxwell construction.
  Bottom: The phase diagram of the annealed model at zero temperature
  in the $(c,\alpha)$ plane.}
\label{fig:fig2}
\end{figure}

By taking the limit $\beta \to \infty$ we have
\begin{equation}
  f_{\alpha,\infty,c}(r) = -r \log r -(1-r) \log (1-r)
  -\frac{c}{\alpha} r + \frac{1}{\alpha} \log(I_0(cr))\:.
\end{equation}
The self-consistent equation now reads (noticing that $I_0'=I_1$)
\begin{equation}
r  = (1-r) e^{\frac{c}{\alpha}(I_1/I_0 -1)}\;,
\end{equation}
and can be simply solved in $\alpha$. In this way we obtain the $T=0$
equation of state
\begin{equation}
  \label{eq:eqstate}
  \alpha(r) = c \frac{1-I_1(cr)/I_0(cr)}{\log(1/r-1)}\;.
\end{equation}
 In Fig. \ref{fig:fig2} we can see that for some values of $\alpha,c$
there are three solutions (one on an unstable branch): the curves
(\ref{eq:eqstate}) are similar to the isothermal curves of a Van Der
Waals fluid and the equilibrium curve can be obtained by Maxwell construction (Fig. \ref{fig:fig2}, right).
We can calculate the stationary points from $\alpha'(r)=0$, whose
solutions inserted back into (\ref{eq:eqstate}) give a coexistence
region that shrinks into a second-order critical point.  The situation
is summarized in the phase diagram in Fig. \ref{fig:fig2} (bottom): there are
two phases with respectively low (UNSAT) and high (SAT) values of $r$ with a
coexistence region that shrinks into a second order critical point
akin to the Van Der Waals phase diagram of a fluid.

\subsection{The annealed approximation for $n_i=0,\dots,L$, $T=0$.}
Consider the annealed sum, where now $n_j=0,1,\dots,L$:
\begin{eqnarray}
\overline{Z} = \sum_{\{n_j\}} \overline{e^{-\beta H}} = \nonumber\\
                  =\sum_{\{n_j\}} \prod_\mu \int dm_\mu e^{-\beta m_\mu^2} \overline{\delta(m_\mu - \sum_j \xi_{j \mu} n_j)} = \nonumber\\
                  = \bigg(\frac{1}{2\pi}\bigg)^{M}\sum_{\{n_j\}} \int \int d \vec{\lambda}  d \vec{m}  e^{-\beta |\vec{m}|^2 + i \vec{\lambda} \cdot \vec{m}} \prod_{j\mu}  \overline{e^{-i n_j \lambda_\mu \xi_{j \mu} }} =\nonumber\\
                  = \bigg(\frac{1}{4\pi\beta}\bigg)^{\frac{M}{2}}\int d \vec{\lambda}   e^{-\frac{1}{4\beta} |\vec{\lambda}|^2} \sum_{\{n_j\}} \prod_{j\mu}(1-p + p \cos(\lambda_\mu n_j)) = \nonumber \\
                  = \bigg(\frac{1}{4\pi\beta}\bigg)^{\frac{M}{2}} \int d \vec{\lambda}  e^{-\frac{1}{4\beta} |\vec{\lambda}|^2} (1+\sum_{s=1}^L \prod_\mu (1-p + p \cos(s \lambda_\mu)))^N  \nonumber.                   
\end{eqnarray} 
We can expand 
\begin{eqnarray}
(1+\sum_{s=1}^L \prod_\mu (1-p + p \cos(s \lambda_\mu)))^N = \nonumber \\
\sum_{k_1+\dots k_L\leq N} {N \choose k_1 \dots k_L} \frac{1}{(N-\sum_i k_i)!}(1-p)^{M\sum_i k_i}\prod_{s=1}^L \prod_\mu (1 + \frac{p}{1-p} \cos(s \lambda_\mu))^{k_s} \nonumber
\end{eqnarray}
and we have the integrals
\begin{equation}
(\frac{1}{\sqrt{4\pi\beta}}\int d\lambda e^{-\frac{1}{4\beta} \lambda^2} (1 + \frac{p}{1-p} \cos(s \lambda))^{k_s})^M \nonumber
\end{equation}
that, once again, can be solved upon approximating \\
 $(1 + \frac{p}{1-p} \cos(s \lambda))^{k_s} \simeq e^{k_s x\cos(s\lambda)}$, where $x=\frac{p}{1-p}$, and using the Bessel functions formula 
\begin{equation}
e^{k_sx \cos (s\lambda)} = I_0(k_sx) + 2\sum_n I_n(k_sx) \cos (ns\lambda) \nonumber,
\end{equation}
obtaining
\begin{eqnarray}
\overline{Z} \simeq \nonumber\\
\sum_{k_1+\dots k_L\leq N} {N \choose k_1 \dots k_L} \frac{1}{(N-\sum_i k_i)!}(1-p)^{M\sum_i k_i}\prod_{s=1}^L F_s^M(k_s x) \\ \nonumber
F_s(k_s x) = I_0(k_s x)+2\sum_n I_n(k_s x) e^{-\beta s^2 n^2} \nonumber
\end{eqnarray}
Upon defining the variables $r_s=k_s/N$ the saddle point
$\overline{Z}  \simeq e^{N f_L^{SP}}$,
where
\begin{eqnarray}
f_L({\bf r}) = -\sum_s r_s \log(r_s)-(1-\sum_s r_s)\log(1-\sum_s r_s)+\\ \nonumber
+ \frac{1}{\alpha}\sum_s \log(F_s(c r_s ))-\frac{c}{\alpha}\sum_s r_s \\ \nonumber
\end{eqnarray}
it comes from the solution of the following optimization problem:
\begin{eqnarray}
\textrm{Maximize}  & & \quad f_L({\bf r}) \nonumber \\ 
\textrm{Subject to}  & & \quad r_s \geq 0 \nonumber \\ 
& & \quad  \sum_s r_s \leq 1
\end{eqnarray}
\begin{figure}[h!!!!]
\centering
\includegraphics*[width=.7\textwidth,angle=0]{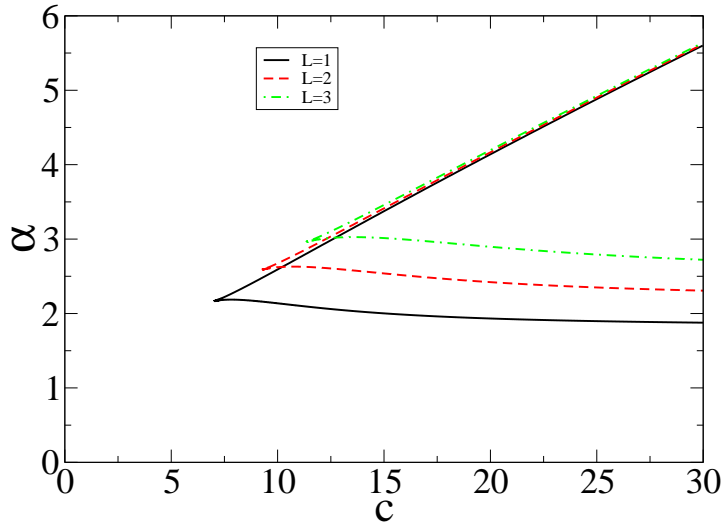}
\caption{Phase diagram in the $(c,\alpha)$ plane from annealed
  calculations for $L=1,2,3$.}
\label{fig:fig3}
\end{figure}
For $\beta \to \infty$, $f_L$ is symmetric under
permutation of the variables. In this limit we search for a symmetric
solution $r_s=r\leq 1/L \quad \forall s$,
and we obtain an extended expression of $f$ for $L>1$:
\begin{equation}
  f_{\alpha,\infty,c,L\geq 1}(r) = -rL \log r -(1-rL) \log (1-rL)
  -\frac{cL}{\alpha} r + \frac{L}{\alpha} \log(I_0(cr))\;.
\end{equation}
Upon taking the maximum, the equation of state now reads
\begin{equation}\label{eqstate}
\alpha(r) = c \frac{1-I_1(cr)/I_0(cr)}{\log(1/r-L)}.
\end{equation}
The picture is qualitatively the same as before, i.e. a Van der Waals picture with first order transition ending in a second order critical point. The values of the first order critical and spinodal lines shift to higher values of $c$ at increasing $L$. In fig \ref{fig:fig3}  we plot the boundary off the coexistence region in the plane $(c,\alpha)$ for several values of $L=1,2,3$: upon increasing $L$ we witness simply to a shift of such a region towards higher values of $c$ and $\alpha$.

\subsection{The quenched case for $n_i=0,1$, $T=0$}
Let us now look to 
the quenched case, where we assume that
the coefficients  $\xi_{j,\mu}$ are fixed, and compute the expectation
value of the logarithm of the partition function $Z$.
We will use replicas, and by introducing $n$ copies
of the system labeled by a further index $a$, $\{n_{j,a}\}$, we will
compute
\begin{eqnarray}
  \overline{Z^n} = \sum_{\{n_{j,a}\}}
\overline{\prod_{a,\mu} e^{-\beta\left(\sum_j \xi_{j,\mu} n_{j,a}\right)^2}} \nonumber\\
  = \sum_{\{n_{j,a}\}} \left( \frac{1}{4 \pi \beta}
  \right)^{\frac{Mn}{2}}
  \int \left(\prod_{a,\mu} d\lambda_{\mu,a} e^{-\frac{ \lambda^2_{\mu,a}}{4\beta}}\right)
  \prod_{j,\mu} \overline{e^{-i\xi_{j,\mu}
    \sum_a n_{j,a}\lambda_{\mu,a} }} \nonumber\\
  = \left( \frac{1}{4 \pi \beta} \right)^{\frac{Mn}{2}}
  \int \left(\prod_{a,\mu} d\lambda_{\mu,a}
  e^{-\frac{ \lambda^2_{\mu,a}}{4\beta}}\right) \sum_{\{n_{j,a}\}}
  \prod_{j,\mu} \left(1-p+p\cos\left(\sum_a n_{j,a}\lambda_{\mu,a}\right)\right) \nonumber\\
  = \left( \frac{1}{4 \pi \beta} \right)^{\frac{Mn}{2}}
  \int \left( \prod_{a,\mu}d\lambda_{\mu,a}
  e^{-\frac{ \lambda^2_{\mu,a}}{4\beta} }\right)
  \left(\sum_{\{n_a\}} \prod_{\mu}
  \left(1-p+p\cos\left(\sum_a n_{a}\lambda_{\mu,a}\right)\right)\right)^N \nonumber \\
  \simeq \sum_{k_1,\dots k_n,k_{12},\dots k_{1\dots n}: \sum k\leq N}
         {N \choose k_1\dots k_{1\dots n} }
         \frac{1}{\left(N-\sum k\right)!} \left(1-p\right)^{M\sum k}
         \\
         \nonumber 
         \left(  \left( \frac{1}{4 \pi \beta} \right)^{\frac{n}{2}}
         \int   \left(\prod_{a} d\lambda_a e^{-\frac{\lambda^2_a}{4\beta}}\right)
         e^{ k_1 x \cos\left(\lambda_1\right) + \dots + k_{12} x
           \cos\left(\lambda_1+\lambda_2\right)
           +\dots +k_{1\dots n} \cos\left(\sum_{a=1,n} \lambda_a\right)}\right)^M\;.   
\end{eqnarray}
By expanding, as before, with first order modified Bessel functions we
get that
\begin{eqnarray}
  e^{ k_1 x \cos\left(\lambda_1\right) +
    \dots + k_{12} x \cos\left(\lambda_1+\lambda_2\right)
    +\dots +k_{1\dots n} \cos\left(\sum_{a=1,n} \lambda_a\right)}= \nonumber\\
  = \left(I_0\left(k_1 x\right)+2\sum_m I_m\left(k_1 x\right)
  \cos\left(m\lambda_1\right)\right)\dots\nonumber\\
  \left(I_0\left(k_{1\dots r} x\right)+2\sum_m
  I_m\left(k_{1\dots n} x\right)
  \cos\left(m\sum_a\lambda_a\right)\right)
  \nonumber\\
=I_0\left(k_1 x\right)I_0\left(k_2 x\right)\cdots I_0\left(k_{1\dots
  n}x\right)
+\mbox{  terms containing at least a cosine.  }
\end{eqnarray}
The terms containing at least a cosine function
upon integration produce terms proportional to $e^{-\beta}$,
that go to zero
when $\beta \to \infty$, that is in the limit of interest for us.
So we have that
\begin{figure}[ht]
\centering
\includegraphics*[width=.75\textwidth,angle=0]{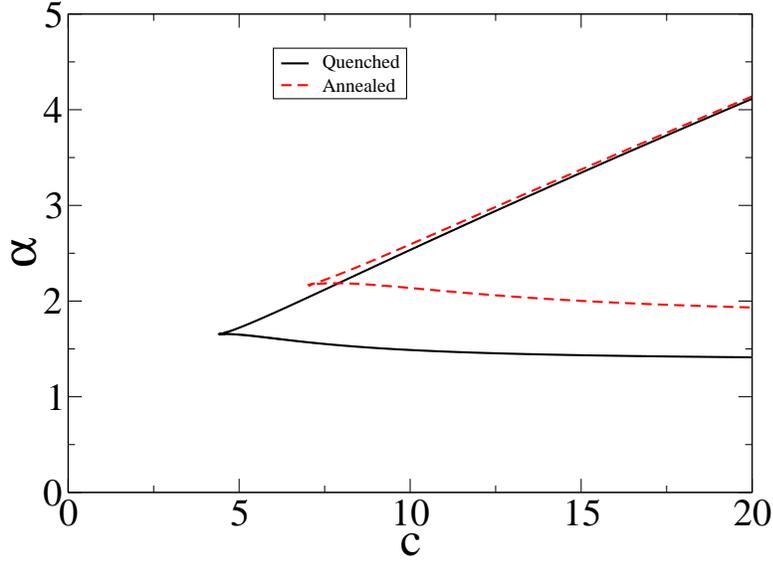}
\caption{Boundary of the coexistence region in the $(c,\alpha)$ plane
  from quenched and annealed calculations.}
\label{fig:fig5}
\end{figure}
\begin{eqnarray}
\lim_{T \to 0}\overline{ Z^n } =
\sum_{k_1,\dots k_n,k_{12},\dots k_{1\dots n}: \sum k\leq N}
    {N \choose k_1 \dots k_{1\dots n} } \frac{1}{\left(N-\sum
      k\right)!}
    \left(1-p\right)^{M\sum k } \nonumber \\ 
    \left(I_0\left(k_1 x\right)I_0\left(k_2 x\right)\cdots I_0
    \left(k_{1\dots n}x\right)\right)^M \simeq
    \int_{r_t\geq 0, \sum_t r_t\leq 1} d {\bf r}\;
    e^{N \; F_{c,\alpha}\left({\bf r}\right)}\;,
\end{eqnarray}
where $t$ runs over the $2^n-1$ possible combinations of $n$ indices.
This defines, with the same set of approximation we had used before
in the annealed case, 
\begin{eqnarray}
  F_{c,\alpha}\left({\bf r}\right)
  = -\sum_t r_t \log\left(r_t\right)
  -\left(1-\sum_t r_t\right)\log\left(1-\sum_t r_t\right)+ \nonumber \\
  + \frac{1}{\alpha}\sum_t \log\left(I_0\left(c r_t \right)\right)
  -\frac{c}{\alpha}\sum_t r_t\;.
\end{eqnarray}
The saddle point approximation can be computed by solving the
optimization problem (with $t= \{ 1\dots 2^n-1 \}$) where one
maximizes $F_{c,\alpha}({\bf r})$ under the constraints $r_t \geq 0$
and $\sum_t r_t \leq 1$.  Both the function $F_{c,\alpha}({\bf r})$
and the domain of the search are symmetric under permutation of
variables. We look for a replica symmetric solution, with $r_t =r$.
The case $n=1$ coincides with the annealed approximation. In the limit
$n \to 0$ we have:
$$
  f_{RS}\left(r\right) = \lim_{n \to 0} F\left(r\right)/n =
  \left( -r\left(\log\left(r\right)-1 \right)
  +\frac{1}{\alpha} \log\left(I_0\left(cr\right)\right)
  -\frac{c}{\alpha} r\right)\log 2\;. 
$$
The equation of state $\frac{\partial f_{RS}}{\partial r}=0$ is
\begin{equation}
  \alpha = c\; \frac{1-\frac{I_1\left(cr\right)}
    {I_0\left(cr\right)}}{\log\left(\frac{1}{r}\right)}\;.
\end{equation}
Interestingly, in this case the boundary of the coexistence
region can be given in parametric form from the equation
$\frac{\partial \alpha(r)}{\partial r}=0$, i.e.:
\begin{equation}
  \log\left(1/r\right) = \frac{1-\frac{I_1\left(x\right)}
    {I_0\left(x\right)}}{x\left(1/2+1/2
    \frac{I_2\left(x\right)}{I_0\left(x\right)}
    -\left(\frac{I_1\left(x\right)}{I_0\left(x\right)}\right)^2\right)}\;,
\end{equation}
where $x \equiv c\,r$.

Figure \ref{fig:fig5} shows a behavior that is qualitatively very
similar to the one of the annealed approximation (see
Fig. \ref{fig:fig2}).  Upon looking at the shrinking of the
coexistence region we can see that $\delta r \propto (c-c_c)^{1/2}$,
i.e. it has a mean field behavior.

We will show how the knowledge of the system thermodynamics
discussed so far can be used to analyze algorithms solving
integer-linear optimization problems. We will show that the difficulty
of performing an optimization task is related to the position of the
instance in the phase diagram. In essence, sticking to the
Van-Der-Waals metaphor, we can distinguish a purely gaseous ($c<c_c$)
from a vapor ($c>c_c$) phase, where the finite size scaling exponent
of the execution time increases and becomes strongly
parameters-dependent, in particular reaching a maximum in the
coexistence region in correspondence of the transition line.

\section{Resolution of single instances}
\label{single}
The statistical mechanical analysis of the system, that we have
discussed in the previous section, is of paramount importance, since
it allows us to qualify the typical behavior of the system, to discuss
fluctuations, to examine critical and collective behavior and phase
transitions.  A second side of the problem is very important, and it
is based on analyzing single instances of the problem. We do not
consider here statistical averages, but the detailed behavior of a
system for a given realization of the coefficients. For example the
work of \cite{mezard2002analytic} has stressed how useful this can be
(in the context, in that case, of satisfiability problems), even for
developing new, powerful optimization algorithms.  We consider here
for simplicity a linear optimization problem (integer programming)
defined over the system (1), where we maximize the objective function
\begin{equation}
f=\frac{1}{N}\sum_i n_i.
\end{equation}
We employ the Matlab\textsuperscript{\textcopyright} solver
\texttt{intlinprog}. We consider two vertical lines in the phase
diagram at fixed $c=4, 12$ (on different sides with respect to the
critical point $c_c$) and perform a sweep over $\alpha \in \left[
  1,5\right]$, $20$ points with fixed number of variables $N=50$
averaged over $10^3$ instances.  In the figure below we report the
relative average length of solutions (left) and the typical machine
time (right, quad-core running at $2.7$ GHz) as a function of $\alpha$
for the two cases.
\begin{figure}[h!!!!]
\centering
\includegraphics*[width=.45\textwidth,angle=0]{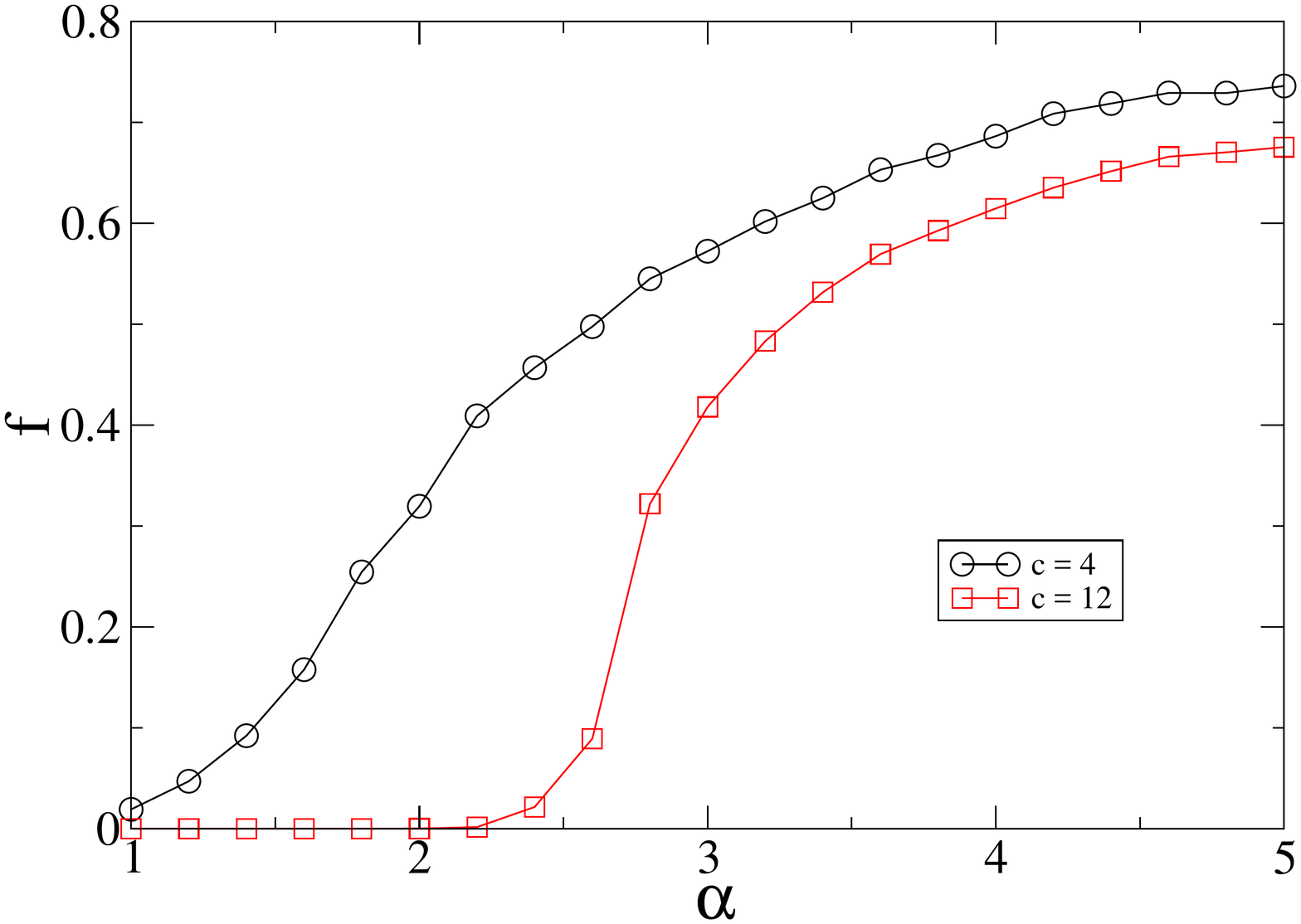}
\includegraphics*[width=.45\textwidth,angle=0]{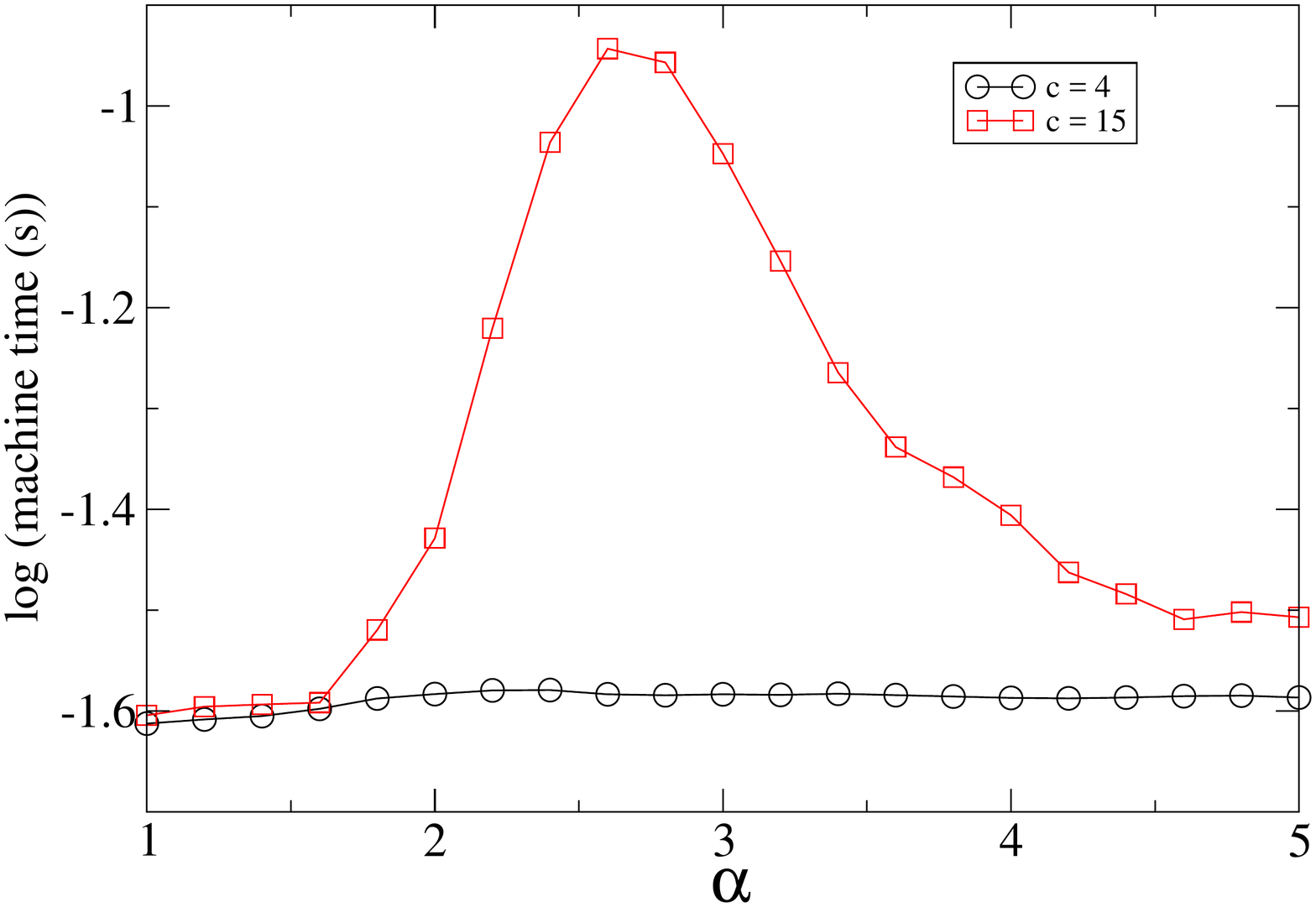}
\caption{Integer programming over random instances with $N=50$ variables. Average  value of the objective function  $f$ (left) and the typical machine time (right) as a function of $\alpha$ for $c=4,12$}
\end{figure}
In both cases we see a crossover from an UNSAT to a SAT phase that
seems to be steeper for $c=12$. The machine time is almost constant
for $c=4$ while for $c=12$ it develops a maximum in correspondence of
the coexistence region.  Next we studied the machine time
as a function of the system size ($N$ number of variables) upon fixing
the point in the phase diagram $(c,\alpha)$ (finite size scaling): on
two vertical lines ($c=4,12$) we consider $\alpha=1.5, 2.5, 4$.
\begin{figure}[h!!!!]
\centering
\includegraphics*[width=.45\textwidth,angle=0]{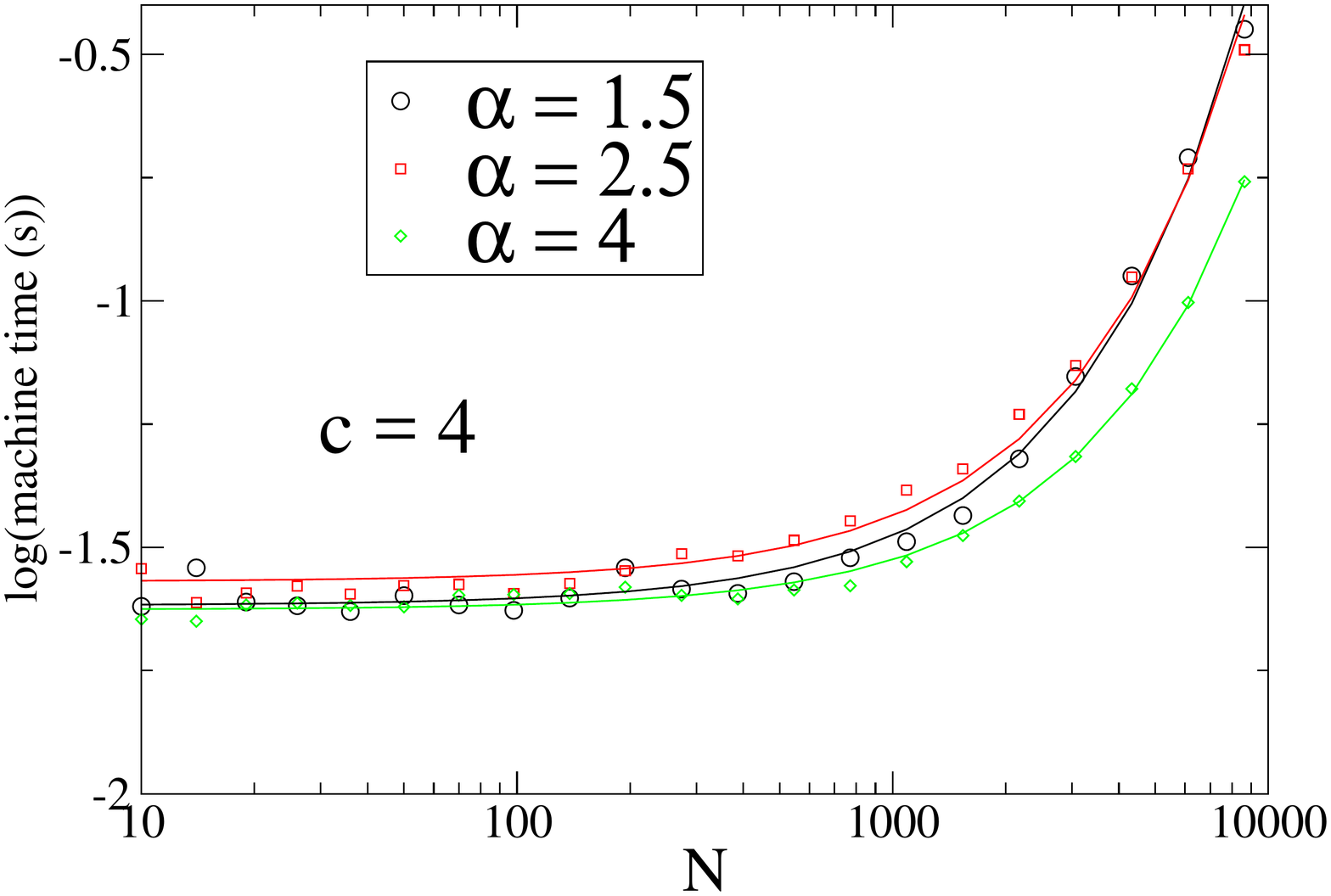}
\includegraphics*[width=.45\textwidth,angle=0]{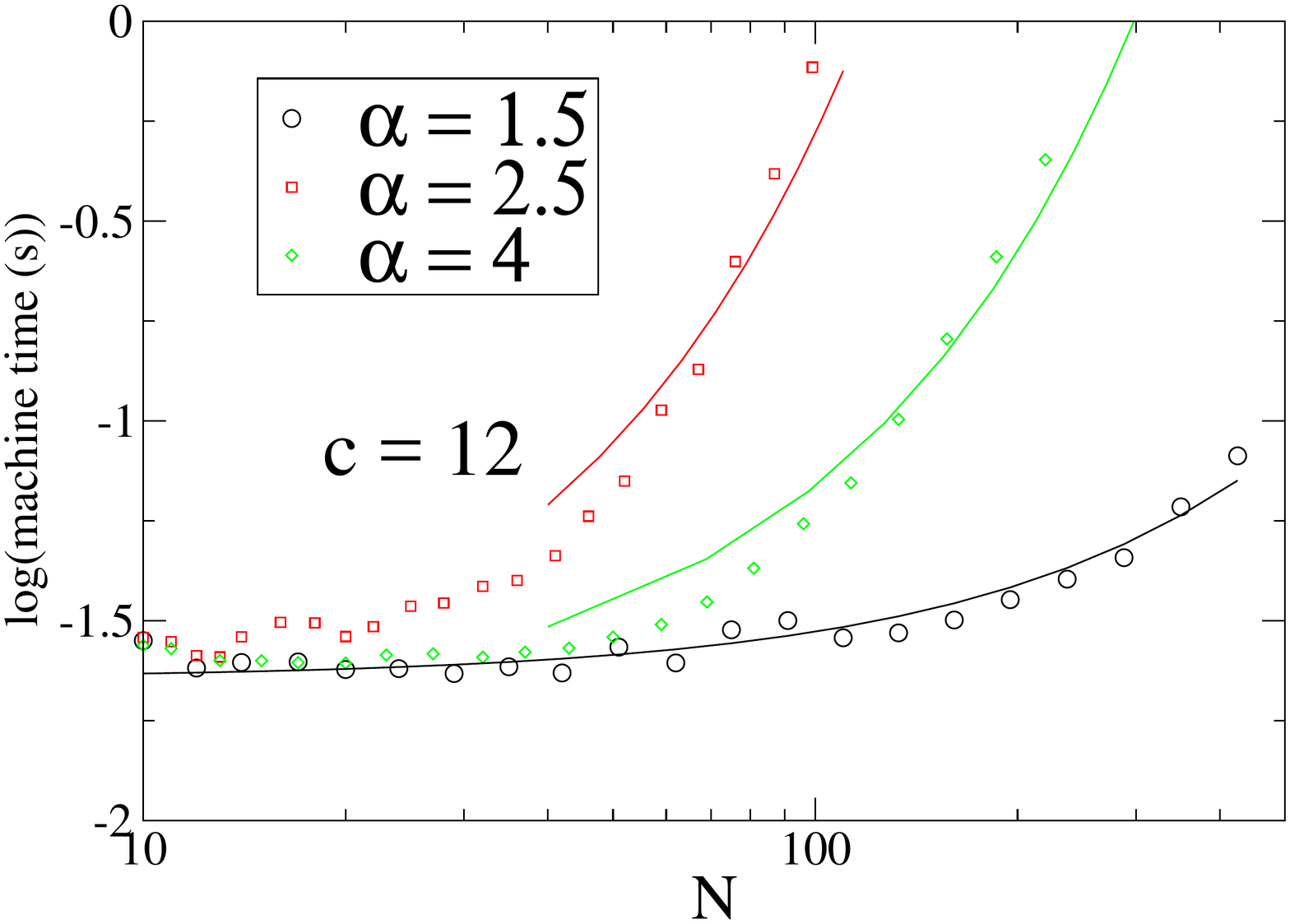}
\caption{Integer programming over random instances. Machine time as a function of the size ($N$ number of variables) for $c=4$ (left), $c=12$ (right) respectively for $\alpha=1.5, 2.5, 4$ }
\end{figure}
In all cases we observe an exponential trend $t=t_0 e^{B N}$. While for
$c=4$ the exponent is small ($B\simeq 10^{-4}$, i.e. we arrive up to
$N=10^4$) and weakly dependent on $\alpha$, for $c=12$ the exponent is
big ($B\simeq 10^{-2}$ we arrive up to $N\simeq O(10^2)$) and
dependent on $\alpha$, with a peak in the coexistence region.

\section*{Conclusions}
The analysis and resolution of difficult combinatorial problems has
gained many insights from statistical mechanics, both in terms of a
characterization of general properties of the solution space and in
terms of providing new powerful resolution algorithms.  A widely
studied class of hard problems is the class of integer linear systems,
given the wide scope of applications and the theoretical insights that
are inherited from linear and convex algebra.  In this note we have
considered ensembles of sparse uncorrelated random systems
parametrised by the average number $c$ of variables per equation and
the ratio $\alpha=N/M$ between total number of variables $N$ and total
number of constraints $M$. We have shown that the space of feasible
solutions endows a Van-Der-Waals phase diagram in the plane ($c$,
$\alpha$).  We have found that the phase transition scenario
depends on the integrality constraint, eventually vanishing for
generic positive integers.  We gave numerical evidence that associated
computational problems become more difficult in regions where a
SAT-UNSAT phase transition is present and, in particular, in the
coexistence region.
 
Several further analysis could be of interest. First, we
gave a picture for the whole solution space while it would be
interesting to consider specific sets of optimization problems upon
adding additional terms in the Hamiltonian (\ref{eq:statmech}).
It is worthwhile
noticing that the phase transition scenario that we have depicted is
essentially within the universality class of the Ising model at odds
with K-SAT problems that endows more complex phase transition scenario
\cite{crisanti2002the}. Along this line a possible development would
be to test the robustness of the universality class across different
ensembles in order to check whether it is related to some universal
feature of the problem and at odds with the K-SAT behavior, e.g. in
the convexity property.  Finally, this statistical mechanical
treatment could give relevant insights in algorithmic strategies like
message passing and Monte Carlo, whose application in integer linear
problems has been already used in a different context\cite{de2014identifying}.

\section*{Appendix: Modified Bessel function of the first kind} 
We report here some useful formulas on modified Bessel function of the
first kind from\cite{abramowitz1964handbook} that have been used in
the above calculations (analytical and/or
numerical).\\
Representations:
$$
I_n(z) = \frac{1}{\pi} \int_{0}^{\pi} d \theta e^{z \cos\theta}\cos(n\theta) = (\frac{1}{2} z^2)^n \sum_k \frac{(\frac{z^2}{4})^k}{k!(n+k)!}
$$
Generating function:
$$
e^{x \cos \lambda} = I_0(x) + 2\sum_n I_n(x) \cos (n\lambda)
$$
Approximation:
\begin{eqnarray}
I_0(x)e^{-x}/\sqrt{x} = \nonumber \\ \nonumber
=1 +3.51562~t+3.08994~t^2+1.2067492~t^3+\\ \nonumber
0.26597~t^4+0.0360768~t^5+0.00458~t^6  \\ \nonumber
\quad t=(x/3.75)^2 \mbox{ for } \quad x<3.75\;,\\ \nonumber
= 0.39894228+0.01328592/t+0.00225/t^2-0.00157565/t^3+\\ \nonumber
0.00916281/t^4-0.02057706/t^5+0.02635537/t^6-\\
0.01647633/t^7+0.00392377/t^8\;,\nonumber\\ 
\quad t = x/3.75 \mbox{ for } \quad x \geq 3.75
\nonumber
\end{eqnarray} 
Derivatives:
\begin{eqnarray} \nonumber
I_0'(x) =I_1(x) \\ \nonumber
I_n'(x) =\frac{1}{2}(I_{n-1}(x) + I_{n+1}(x))
\end{eqnarray}
For ratios see\cite{amos1974computation}:
\begin{eqnarray} \nonumber
r_n =\frac{I_{n+1}}{I_n} \\ \nonumber
r_n^0= \frac{x}{n+1/2+((n+3/2)^2+x^2)^{1/2}}\\ \nonumber
R_{n+1}^m = \frac{r_{n+1}^m}{r_n^m} \\ \nonumber
r_n^{m+1} = \frac{x}{n+1+((n+1)^2+x^2 R_{n+1}^m)^{1/2}}
\end{eqnarray}

\section*{Acknowledgments}
The research leading to these results has received funding from the People Programme (Marie Curie Actions) of the European Union's Seventh Framework Programme ($FP7/2007-2013$) under REA grant agreement $n[291734]$ (DDM).
DDM would like to thank Alexandr Kazda and Davide Raimondo for interesting discussions. 
\section*{References}
\bibliographystyle{unsrt}
\bibliography{ILref}
\end{document}